\documentstyle[12pt,aasms4]{article}

\def\be{\begin{equation}}
\def\ee{\end{equation}}
\def\ba{\begin{eqnarray}}
\def\ea{\end{eqnarray}}

\def\Rbar{\overline{R}}

\def\sigmab{\overline{\sigma}}
\def\ybar{\overline{y}}

\def\la{\mathrel{\mathpalette\fun <}}
\def\ga{\mathrel{\mathpalette\fun >}}
\def\fun#1#2{\lower3.6pt\vbox{\baselineskip0pt\lineskip.9pt
        \ialign{$\mathsurround=0pt#1\hfill##\hfil$\crcr#2\crcr\sim\crcr}}}

\slugcomment{To be submitted to ApJ}

\begin{document}
\null\vspace{-62pt}
\begin{flushright}
FERMILAB-Pub-96/060-A\\
%version date: March 8, 1996\\
\today
\end{flushright}

\title{Statistics of Extreme Gravitational Lensing Events. II.\\
The Small Shear Case}

\author{Yun Wang}
\affil{{\it NASA/Fermilab Astrophysics Center} \\
{\it Fermi National Accelerator Laboratory, Batavia, IL 60510-0500\\}
{\it email: ywang@fnas12.fnal.gov}}

\author{Edwin L. Turner}
\affil{{\it Princeton University Observatory} \\
{\it Peyton Hall, Princeton, NJ 08544\\}
{\it email: elt@astro.princeton.edu}}

\vspace{.4in}
\centerline{\bf Abstract}
\begin{quotation}

We consider an astrophysical system with a population of sources
and a population of lenses. For each pair of source and lens,
there is a thin on-axis tube-like volume behind the lens 
in which the radiation flux from the source 
is greatly increased due to gravitational lensing.
Any objects (such as dust grains) which pass through such a thin tube
will experience strong bursts of radiation, i.e.,  
Extreme Gravitational Lensing Events (EGLE).
We study the physics and statistics of EGLE for the case in which 
the shear is larger or comparable to the finite source size.
We find that the presence of shear has only a small effect on the
EGLE statistics.

\end{quotation}

%\keywords{gravitational lensing}

\section{Introduction}

Gravitational lensing in the context of direct observations of lensed
sources has been studied extensively in past years (\cite{BNrev92},
\cite{Book92} and references therein). Recently, we proposed a new way of looking at
gravitational lensing, the statistics of extreme gravitational lensing
events (EGLE), which may have significant astrophysical effects
(\cite{Wang96}, hereafter referred to as Paper I).
We introduced the basics of EGLE in Paper I, where we considered 
the case in which the finite source size is more important than shear.
In this paper, we discuss the case in which shear can not be neglected.

Extreme gravitational lensing events (EGLE) occur when the source,
lens, and observer (or target) are nearly on-axis.
In the statistics of EGLE,
we consider an astrophysical system of diameter $D_{\rm c}$, in
which both the sources and lenses are uniformly distributed,
and the targets can be the fragile objects in the same system,
such as dust grains, etc.

If the target moves a distance $d$ away from the line 
connecting the source and the lens, it is equivalent to the source 
moving an angular distance of $y$ from the optical axis (the line connecting 
the lens and the target). Measuring $y$ in units of the angular 
Einstein radius, we have
\be
\label{eq:O-S}
y \simeq \left( \frac{D_{\rm ds}}{D_{\rm d}} \right)\,
\frac{d}{D_{\rm s} \theta_{\rm E}},
\ee
where $D_{\rm ds}$, $D_{\rm s}$, and $D_{\rm d}$ are angular 
diameter distances between the lens and source, target and source, 
target and lens respectively. $\theta_{\rm E}= \sqrt{ 2 R_{\rm S} 
D_{\rm ds}/(D_{\rm d} D_{\rm s})}$ is the angular Einstein radius.  
$R_{\rm S}=2GM$ is the Schwarzschild radius of the lens with mass $M$. 
Eq.(\ref{eq:O-S}) is the small angle approximation, 
valid for $d \ll D_{\rm d}$. The dimensionless radius of a source
with physical radius $\rho$ is defined to be $R\equiv \rho/(D_{\rm s}
\theta_{\rm E})$.

For a given pair of lens and source, the shear $\gamma$ 
due to other mass inhomogeneities near the line-of-sight 
is the same order of magnitude as the 
optical depth for microlensing, $\tau$, the probability that the 
source is lensed. We find [see Appendix]
\be
\label{eq:gamma}
\gamma = \sqrt{2}\,\, \zeta\left(\frac{3}{2}\right)\, \tau
\simeq 3.7\, \tau.
\ee
$\zeta\left(x\right)$ is the Riemann zeta function.
Not surprisingly, the statistics of EGLE is much more complicated for 
sheared lenses than for isolated lenses ($\gamma=0$). 

The lens model for a perturbed Schwarzschild lens (i.e., point
mass lens with shear) has been investigated by 
Chang and Refsdal (1979, 1984) and Subramanian and Chitre (1985).
The magnification of extended sources by a perturbed Schwarzschild lens
has been studied by Chang (1984), Schneider and Weiss (1987).
In Paper I, we showed that shear is not important for $\gamma <R \ll 1$. 

In this paper, we study the statistics of EGLE for sources with 
small dimensionless radius $R$ and slightly sheared lenses ($\gamma\ll 1$),
with $\gamma \geq R$. 
We generally follow the notation and conventions of Schneider et al. (1992)
and Paper I.

\section{The perturbed Schwarzschild lens}

In this section, we study the basic properties of a 
perturbed Schwarzschild lens.
In the absence of shear, the caustic (which corresponds to infinite 
magnification) of a point mass lens in the source plane is a point 
(intersection with the optical axis). 
If the shear of a macrolens at the location of the point mass lens is 
$ 0<\gamma<1$, the caustic in the source plane changes from a point 
to an astroid shaped closed curve with four cusps, 
assuming that the surface mass density
of the macrolens is zero at the location of the point mass lens.

Let us define
\be
{\bf Y}= \frac{{\bf y}}{\sqrt{1+\gamma}}, \hskip 1cm
\Lambda = \frac{1-\gamma}{1+\gamma}.
\ee
The caustic curve intersects the $Y_1$ axis at $\pm (1-\Lambda)$,
the $Y_2$ axis at $\pm (1-\Lambda)/\sqrt{\Lambda}$. For $\gamma \ll 1$,
the caustic is described by
\be
Y_1 = -2\gamma \cos^3\phi, \hskip 1cm Y_2 = 2\gamma \sin^3\phi.
\ee
As we discussed before, moving the observer away from the line
connecting the source and lens is equivalent to moving the source 
relative to the optical axis.
When a source moves off the optical axis, if it moves along 
$Y_1=0$ or $Y_2=0$, it crosses a cusp; if it moves along $Y_1=Y_2$, 
it crosses a fold. All observables for a source moving off the optical
axis in an arbitrary direction fall between the two limits set by 
$Y_1Y_2=0$ and $Y_1=Y_2$. 
We therefore restrict our attention to these two relatively simple
cases below.

The lens mapping equation can be solved exactly for $Y_1Y_2=0$.
(Chang \& Refsdal 1979 \& 1984, Subramanian \& Chitre 1985)
For a point source moving along the $Y_1$ axis, its magnification is
\ba
&& \mu_{\rm p}\left(|Y_1|\leq 1-\Lambda\right) = \frac{1}{(1+\gamma)^2}
\,\frac{ (1-\Lambda) (1+\Lambda) -Y_1^2}{\Lambda \left[ (1-\Lambda)^2-Y_1^2\right]}, \nonumber\\
&& \mu_{\rm p}\left(|Y_1|>1-\Lambda\right) = \frac{1}{(1+\gamma)^2}\,
\frac{ Y_1\left[ Y_1^2 +(3\Lambda-1)\right]}{\Lambda \sqrt{Y_1^2+4\Lambda}
\left[Y_1^2- (1-\Lambda)^2 \right]}.
\ea
For a point source moving along the $Y_2$ axis, its magnification is
\ba
&& \mu_{\rm p} \left(|Y_2|\leq \frac{1-\Lambda}{\sqrt{\Lambda}}\right) 
= \frac{1}{(1+\gamma)^2}\,\frac{ (1-\Lambda)(1+\Lambda) +\Lambda Y_2^2}
{\Lambda \left[ (1-\Lambda)^2-\Lambda Y_2^2\right]}, \nonumber\\
&& \mu_{\rm p}\left(|Y_2|> \frac{1-\Lambda}{\sqrt{\Lambda}}\right) 
= \frac{1}{(1+\gamma)^2}\,\frac{ Y_2\left[ Y_2^2 +(3-\Lambda)\right]}
{\sqrt{Y_2^2+4} \left[\Lambda Y_2^2- (1-\Lambda)^2 \right]}.
\ea

For $\gamma\ll 1$, $1-\Lambda \simeq 2\gamma$, ${\bf Y} \simeq {\bf y}$. 
Let us define
\be
\overline{\bf y}= \frac{{\bf y}}{2\gamma}, \hskip 1cm
\overline{ R}= \frac{R}{2\gamma}.
\ee
Then, for $y_2=0$,
\be
\label{eq:mu_cusp}
\mu_{\rm p}\left(|\overline{y}_1|\leq 1\right) = \frac{1}{\gamma}\, 
\frac{1}{1-\overline{y}_1^2}, \hskip 1cm
\mu_{\rm p}\left(|\overline{y}_1|> 1\right)= \frac{1}{\gamma}\, 
\frac{\overline{y}_1}{2\left(\overline{y}_1^2-1\right)}
\, \frac{1+2\gamma^2 \overline{y}_1^2}{\sqrt{1+\gamma^2\overline{y}_1^2}}.
\ee
For a small source with $\overline{R} \la 0.2$, the maximum magnification is
given by $\mu_{\rm p}\left(|\overline{y}_1|\leq 1\right)$, with 
$(1-\overline{y}_1) \simeq \left(1.2\, \overline{R}\right)^{0.73}$.
For $y_1=0$, just replace $y_1$ with $y_2$.
Figure 1 shows the lightcurves for $\overline{R}=0.01$, 0.1, 0.4, 1 
and $y_2=0$.
Note that the smaller the source size compared to the shear,
the larger the magnification of the source near the caustic;
the peaks in the lightcurves correspond to caustic-crossing.

For a source moving along the diagonal line $y_1=y_2$,
$(\gamma-y)$ gives the source's distance to the caustic.
For a small source with $\overline{R} \la 0.2$, the maximum magnification is
given by the point source magnification $\mu_{\rm p}(\overline{y})$, 
with $(0.5-\overline{y}) = 0.67\overline{R}$.
The magnification of a point source near the caustic is given by
(\cite{CH84.2})
\be
\mu\left(\overline{y}\rightarrow 0.5\right) \simeq \frac{1}{\gamma}\, 
\frac{0.3}{\sqrt{0.5-\overline{y}}}.
\ee
For a very small source ($\overline{R} \la 0.02$) (\cite{CH84.2}),
\ba
&& \mu_{\rm e}\left(\overline{R}, \overline{y}\right) \simeq 
\frac{1}{\gamma}\, \frac{0.3}{\sqrt{\overline{R}}}\, \zeta(w), 
\hskip 1cm w = \frac{\overline{y}-0.5}{\overline{R}}, \\
&& \zeta(x_0)= \frac{2}{\pi}\int^{z_{\rm max}}_{-1} {\rm d}z\,
\sqrt{ \frac{1-z^2}{-(z+x_0)}}, \hskip 1cm 
{\rm if}\,\,\,|x_0|<1, \,\,\,z_{\rm max}= -x_0; 
\,\,\, {\rm else} \,\,\, z_{\rm max}=1.\nonumber 
\ea
Figure 2 shows the lightcurves for $\overline{R}=0.01$, 0.1, 0.4, 1 
and $y_2=y_1$.
Again, the peaks in the lightcurves correspond to caustic-crossing.

For $\Rbar \ga 0.01$, we fit $\mu_{\rm max}\left(R,\gamma\right)$ 
with the following formulas:
\ba
\label{eq:mu_max}
\gamma \mu_{\rm max} &=& \frac{\left[ 1- w(\Rbar)\right]}{\Rbar}
+\frac{w(\Rbar)}{2 \Rbar^{2/3}} ,  \hskip 1cm
\mbox{cusp-crossing }(y_2=0); \nonumber \\
\gamma \mu_{\rm max} &= & \frac{\left[ 1- w(\Rbar)\right]}{\Rbar}
+\frac{0.3665}{\sqrt{\Rbar}} \,w(\Rbar), \hskip 1cm
\mbox{fold-crossing }(y_2=y_1). 
\ea
$w=\exp\left[-\Rbar(1+\Rbar)/2\right]$ is the weight function.
The above fit formulas underestimate the maximum magnification
by about 20-30\% when the size of the source is comparable to the size 
of the region enclosed by the caustic curve.
Note that for $\gamma=0$ ($\Rbar=\infty$), we recover
$\mu_{\rm max}=2/R$, the exact result used in Paper I.
For $\Rbar<1$ [i.e., $R<2\gamma$], $\mu_{\rm max}<2/R$.

\section{EGLE volume statistics: general description}

Let us consider a mixed population of uniformly distributed 
sources (with number density $n_{\rm S}$) and lenses (with number density 
$n_{\rm L}$) confined in a volume of diameter $D_{\rm c}$.
We are interested in computing the volume fraction of space in which
the flux exceeds $f$ due to gravitational lensing, i.e., the volume
fraction of space in EGLE tubes with flux $>f$.

Note that in EGLE statistics, both $\gamma$ and $R$ depend on distances.
We write
\be
R = R_0 \sqrt{ \frac{t}{x(t+x)} }, \hskip 1cm
\gamma = \left\{\begin{array}{ll}
\gamma_0 \, tx, \hskip 1cm t+x \le 1,\\
\gamma_0 \, tx/(t+x), \hskip 1cm t+x > 1.\end{array}\right.
\label{eq:R,gamma}
\ee
where
\ba
&&t \equiv \frac{D_{\rm d}}{D_{\rm c}}, \hskip 1cm
x \equiv \frac{D_{\rm ds}}{D_{\rm c}}; \nonumber \\
&& R_0 \equiv \frac{\rho}{\sqrt{2R_{\rm S} D_{\rm c}} }, \hskip 1cm
\gamma_0 \equiv 3.7 \tau_0.
\label{eq:define R0,gamma0}
\ea
$\rho$ is the physical radius of the source,
$\tau_0$ is the optical depth at $D_{\rm s}=D_{\rm c}$.

For a source with luminosity $L_{\rm S}$, the unlensed flux $f_0= 
L_{\rm S}/[4\pi D_{\rm s}^2]$. As in Paper I, we define
\be
\alpha \equiv \left.\left[ \frac{\mu_{\rm max}^{\gamma=0}
(R)}{f/f_0}\right]^2
\right|_{D_{\rm s}=D_{\rm c}}^{D_{\rm d}=D_{\rm ds}}
=\left[ \frac{(2/R_0)}{ (f/f_{\rm min})} \right]^2
=\frac{8R_{\rm S} D_{\rm c}}{\rho^2} \left(\frac{L_{\rm S}}
{4\pi D_{\rm c}^2 f}\right)^2,
\ee
where $f_{\rm min}= L_{\rm S}/[4\pi D_{\rm c}^2]$.
$\alpha$ measures the maximum magnification of the source 
relative to the flux $f$, in the limit of zero shear.

We use the simple but reasonable approximation:
\be
\label{eq:app mu_e}
\mu_{\rm e}(y, R, \gamma) \simeq \left\{ \begin{array}{ll}
\mu_{\rm p}(y, \gamma) & \mbox{for $\mu < \mu_{\rm max}$}\\
\,\,\\
\mu_{\rm max} & \mbox{elsewhere} 
\end{array}
\right.
\ee
where $\mu_{\rm p}(y, \gamma)$ is the point source magnification
in the presence of small shear, and $\mu_{\rm max}$ is the maximum 
magnification. For $\gamma=0$, we recover the approximation 
used in Paper I.
Let us write [see Eq.(\ref{eq:mu_max})]
\be
\mu_{\rm max}= \left[ 1- w\left({R}/{\gamma}\right) \right]\,
\frac{2}{R} + w\left({R}/{\gamma}\right) \,\mu_{\rm max}(\gamma\ge R).
\ee
$w\left({R}/{\gamma}\right)$ is a weight function.
For $\gamma \ll R$, we recover the exact formula used in Paper I.

For a given pair of lens and source,
we define the EGLE tube with minimum flux $f$ to be the small tube-shaped
volume behind the lens in which the flux exceeds $f$ due to gravitational
lensing.
For a given pair of lens and source with separation $x=D_{\rm ds}/D_{\rm c}$,
the length of the EGLE tube is $t_{\rm max}=D_{\rm d}^{\rm max}/D_{\rm c}$;
i.e., the EGLE tube extends from $t=0$ to $t=t_{\rm max}$.
The minimum magnification needed at a point inside the EGLE tube is
$\mu =f/f_0 =(t+x)^2\, f/f_{\rm min}$,
the unlensed flux $f_0= L_{\rm S}/[4\pi D_{\rm s}^2]=f_{\rm min}/(t+x)^2 $.
The length of the EGLE tube is given by setting $\mu=\mu_{\rm max}(t,x)$. 

The cross-section of the EGLE tube at a distance of $D_{\rm d}$ behind
the lens along the line SL which connects the source and the lens, 
$\sigma(D_{\rm d}, f)$, is the area perpendicular to SL
in which the magnification exceeds $\mu=f/f_0$, i.e., the flux exceeds $f$.
The cross-section of the EGLE tube can be written in the form
\be
\label{eq:sigma}
\sigma(D_{\rm d}, f) = 2\pi R_{\rm S} D_{\rm c}\,
\left(\frac{f_{\rm min}}{f}\right)^2 \, \left[\frac{t}{x(t+x)^3}\right]\,
\sigma^*, \hskip 1cm
\sigma^* \equiv \frac{\mu^2}{\pi}\, \sigmab.
\ee
$\sigmab =\sigma/[d(\mu)/y(\mu)]^2$ is the cross-section of the EGLE tube 
in units of Einstein radius squared [see Eq.(\ref{eq:O-S})].
For $R=\gamma=0$, $\sigma =\pi d^2(\mu)$, $\mu \simeq 1/y$, hence
$\sigma^*=1$. For $\gamma \neq 0$, $\sigma^* =\sigma^*(\gamma\mu)$.

The volume fraction of space occupied by EGLE tubes in which the flux
exceeds $f$, ${\cal F}_{\rm L}$, can be computed using the same method 
as in Paper I. 

For a given lens and source pair, the associated EGLE tube with minimum
flux $f$ has the volume
\be
V_{\rm SL}(f) = \int_0^{D_{\rm d}^{\rm max}}
 {\rm d} D_{\rm d}\,\sigma(D_{\rm d},f)
= 2\pi R_{\rm S} D_{\rm c}^2 \left(\frac{f_{\rm min}}{f}\right)^2
\cdot \frac{1}{x} \int^{t_{\rm max}}_0 {\rm d}t\,
\frac{t}{(t+x)^3}\, \sigma^*(t,x).
\ee
The volume fraction of space in EGLE tubes of flux $>f$ for a given lens 
is
\be
V_{\rm L}(f)=4 \pi n_{\rm S} \int^{D_{\rm c}}_0 {\rm d} D_{\rm ds}\,
D_{\rm ds}^2 \, V_{\rm SL}(f).
\ee
The volume fraction of space in EGLE tubes of flux $>f$ is
\be
{\cal F}_{\rm L}(f) = n_{\rm L} V_{\rm L}(f)
\equiv {\cal F}_{\rm L}(f,R=\gamma=0)\, I(\alpha, R_0/\gamma_0),
\ee
where ${\cal F}_{\rm L}(f,R=\gamma=0)$ is the volume fraction of space in 
EGLE tubes of flux $>f$ for point source in the absence of shear
(see Paper I),
\be
{\cal F}_{\rm L}(f,R=\gamma=0)= 4\pi^2 n_{\rm S} n_{\rm L} R_{\rm S}
D_{\rm c}^5 \left(\frac{f_{\rm min}}{f}\right)^2,
\ee
and 
\be
I(\alpha, R_0/\gamma_0)= \frac{{\cal F}_{\rm L}(f,\rho, \gamma)}{
{\cal F}_{\rm L}(f,\rho=0=\gamma)}
= 2 \int^1_0 {\rm d}x\, x \int^{t_{\rm max}}_0 {\rm d}t\,
\frac{t}{(t+x)^3}\, \sigma^*(t,x).
\ee
$I(\alpha, R_0/\gamma_0)$ is the modification factor in EGLE 
volume fraction due to non-zero shear or/and finite source size.
For $\gamma_0=0$, $I(\alpha, R_0/\gamma_0)=I(\alpha)$
(see Paper I). 

For given finite ratio $R_0/\gamma_0$, $I(\alpha, R_0/\gamma_0)$ has the
following asymptotic behavior.
At very large $\alpha$, both $R$ and $\gamma$ become negligible, 
and $I(\alpha,R_0/\gamma_0)$ approaches 1.
If $\alpha$ is not too small, EGLE tubes with flux $>f$ exist
for arbitrary source-lens separations ($0<D_{\rm ds}\leq D_{\rm c}$).
When $\alpha$ is sufficiently small, EGLE tubes with flux $>f$ do not
exist for source-lens separations larger than the maximum 
$D_{\rm ds}^{\rm max}< D_{\rm c}$.
For decreasing $\alpha$, the EGLE tubes with flux $>f$ decrease in length 
and they require smaller source-lens separations; when the EGLE tubes
are sufficiently short and require sufficiently small source-lens
separations, $R> \gamma$, the finite size of the source dominates
[see Eq.(\ref{eq:R,gamma})], and $I(\alpha, R_0/\gamma_0)$ approaches
$I(\alpha)$.

\section{EGLE volume statistics for $\gamma \neq 0$, $R\neq 0$}

In the presence of shear, the EGLE tube has very complicated topology.
For given $R$, $\gamma \neq 0$ always leads to
smaller $\mu_{\rm max}$ [see Eq.(\ref{eq:mu_max})], hence shorter EGLE tubes. 
The cross section of an EGLE tube is determined by the shape
of the caustic curve [with both cusps and folds] in the source plane.
For $\gamma \neq 0$, the volume in which the flux
produced by EGLE exceeds the critical value has the topology of
a filled tube over most of its length, but near the end of the tube 
a complex geometry develops reflecting the cusp and fold natures 
of the caustic.

For a point source, the EGLE tube has an astroid shaped cross section until
$D_{\rm d}=D_{\rm d}^{0}$, with $D_{\rm d}^{0}$ given by
$f/f_0(D_{\rm d}^0)=1/\gamma$. For $D_{\rm d}>D_{\rm d}^{0}$,
the EGLE tube is hollow at the center, with walls tapering off to 
infinity away from the center; the walls are thicker and rounded
in the cusp directions, and thinner with steep outside surfaces
half-way between the cusp directions. The walls taper off more quickly 
half-way between the cusp directions than in the cusp directions.

For a finite source, if the source is sufficiently large ($R>2\gamma$),
the EGLE tube ends without the shear-induced hollow and the effect of
shear is negligible. If $R<2\gamma$, the EGLE tube has a topology
similar to the $R=0$ case, except that the shear-induced hollow
has walls which end at a finite distance behind the lens,
the walls stretch furthest in the cusp directions, and shortest
half-way between the cusps; i.e., the end of the EGLE tube is a
hollow volume, with thinning walls which terminate in four rounded
legs centered in the cusp directions.

The volume of an EGLE tube for $1\gg \gamma>R>0$ is between the volume
of an EGLE tube assuming cusp-crossing only and the volume of an EGLE
tube assuming fold-crossing only.
Compared to the point source magnification in the absence of shear
(dashed curve in Figs.1-2), for a given magnification of the source, 
the sheared lens (with $\gamma >R$) leads to a much larger 
cross-section for cusp-crossing [see Figure 1] and a smaller 
cross-section for fold-crossing [see Figure 2].
Therefore, the volume modification factor $I(\alpha,R_0/\gamma_0)$ should
be smaller in the fold-crossing case than in
the zero shear case; but it can be larger in the cusp-crossing case
than in the zero shear case, because of the significant thickening
of the EGLE tubes at the ends away from the lenses. 

In this section, we use Eq.(\ref{eq:mu_max})
to compute the EGLE tube lengths, and $\mu_{\rm p}(y)=\mu_{\rm 
p}(y,\gamma\neq 0, R=0)$ to compute the EGLE tube cross-sections.

The cross-section of the EGLE tube can be expressed in terms of
the dimensionless parameter $\sigma^*=\mu^2\, \sigmab /\pi$
[see Eq.(\ref{eq:sigma})], where
$\sigmab =\sigma/[d(\mu)/y(\mu)]^2$ is the cross-section of the EGLE tube 
in units of Einstein radius squared, 
\be
\sigmab = \pi \left( y_{(2)}^2- y_{(1)}^2 \right).
\ee
$y_{(2)}$ and $y_{(1)}$ are roots of $\mu_e(y,\gamma)=\mu$,
$y_{(2)}>y_{(1)}\geq 0$ (if $\mu_e(y,\gamma)=\mu$ only has one root,
then we denote it by $y_{(2)}$, and set $y_{(1)}=0$).
For $\gamma \ll 1$, $\gamma\,\mu_e(y,\gamma)$ only depends on
$\ybar\equiv y/(2\gamma)$. Then
\be
\sigma^* = 4 \delta^2 \left[ \ybar_{(2)}^2- \ybar_{(1)}^2 \right],
\hskip 1cm \delta \equiv \gamma \mu.
\ee
$\ybar_{(2)}$ and $\ybar_{(1)}$ are roots of $\gamma\,\mu_e(y,\gamma)
\equiv \delta_{\rm e}(\ybar)=\delta$, $y_{(2)}>y_{(1)}\geq 0$ 
($\ybar_{(1)}=0$ if only one root exists).

For $\mu \la 0.1/\gamma$, the shear on the lens has negligible effect
on the lightcurve, hence $\sigma^*=1$ as in the $R=\gamma=0$ case.
For $\mu \ga 0.1/\gamma$, we can compute $\sigma^*$ using the point
source and small shear lightcurves shown in Figs.1-2, for
cusp-crossing and fold-crossing respectively.
For cusp-crossing, we can compute $\sigma^*$ analytically using
Eq.(\ref{eq:mu_cusp}).
We find
\be
\sigma^*(\mbox{cusp-crossing})=\left\{ \begin{array}{ll}
\left[ 1+ \sqrt{1+16 \delta^2}+ 8 \delta \right]/2, \hskip 1cm \delta \ge 1;\\
\left[ 1+ \sqrt{1+16 \delta^2}\right]^2 /4, \hskip 1cm \delta < 1,\end{array}
\right.
\ee
where $\delta \equiv \gamma \mu$. Note that for $\delta =0$ (i.e.,
$\gamma=0$), we recover $\sigma^*=1$.
For fold-crossing, we have to find the roots of $\gamma\,\mu_e(y,\gamma)
\equiv \delta_e(\ybar)=\delta$ numerically to
compute $\sigma^*$.

To estimate the effect of shear on the EGLE volume fractions, 
we write
\be
I_{avg}(R_0, \gamma_0)= \frac{I_{cusp}(R_0, \gamma_0)\, 2\gamma_0 R_0
+I_{fold}(R_0, \gamma_0)\, \left[(2\gamma_0)^2\, A- 2\gamma_0 R_0\right]}
{(2\gamma_0)^2\, A},
\ee
where $A=5\pi/32$ is $\frac{1}{4}$ the area of the astroid enclosed
by the caustic curve [in units of $(2\gamma_0)^2$].
The averaged EGLE volume fraction is only slightly larger than the
EGLE volume fraction assuming fold-crossing only, because most of the
caustic is fold. 

In Fig. 3, we show the modification factor in EGLE volume fraction due to
finite source size and non-zero shear, $I_{avg}(\alpha, R_0/\gamma_0)$,
for $R_0/\gamma_0=10$, 1, 0.1, 0.01, 0.001.
The $R_0/\gamma_0=10$, 1 curves are
indistinguishable from the $\gamma=0$ curve (dashed line).
Note that all curves in Fig.3 converge to the $\gamma=0$ curve
at small and large $\alpha$, as discussed at the end of last section.
Note that the effect of $\gamma \neq 0$ decreases at both extremes
of large $\alpha$ [small flux or high magnification]
and small $\alpha$ [high flux or low magnification].
Clearly, the presence of shear has only a moderate
effect on the EGLE volume fractions. 

For objects which enter an EGLE tube perpendicular to the lens-source line,
the EGLE durations are longest in the cusp directions, and shortest 
half-way between the cusps. Since the EGLE durations also depend on
the angle at which an object enters the EGLE tube, the effect of shear
on the EGLE durations should be negligible statistically also.

\section{Summary}

We find that the presence of small shear [comparable or larger
than the dimensionless source size] has only a small effect on the
EGLE volume fractions. 
Specifically, the EGLE volumes are typically decreased
by factors of a few for $\gamma_0 \sim 10^2 R_0$
[the characteristic dimensionless shear $\gamma_0$ and 
source size $R_0$ are
defined in Eq.(\ref{eq:define R0,gamma0})]
at moderate values of $\alpha$ [which measures the 
maximum magnification of the source in the limit of zero shear
relative to the flux $f$] and otherwise unaffected,
see Fig. 3. This means that the $\gamma=0$ results given in
Paper I will be adequate for most order of magnitude considerations.
 
It is quite reasonable that the effect of $R \la \gamma
\ll 1$ is modest since the convergence provided by the
EGLE lens itself constrains the total flux concentrated into
the EGLE tube. The small shear considered in this paper only
rearranges this flux somewhat into a slightly modified EGLE volume,
especially near the end of the EGLE tube.

Y.W. is supported by the DOE and NASA under Grant NAG5-2788.
E.L.T. gratefully acknowledges support from NSF grant AST94-19400.

\vskip 1in

\appendix{\bf Appendix: Derivation of $\gamma(\tau)$}

Here we derive the shear $\gamma$ 
due to other mass inhomogeneities near the line-of-sight 
for a given pair of lens and source, in terms of 
the optical depth for microlensing, $\tau$, the probability that the 
source is lensed. 

We assume that the EGLE lens 
[the lens under consideration] is embedded in a distribution of 
identical lenses with surface number density $\Sigma$,
each lens has the same angular Einstein radius $\theta_E$
associated with it.
The total shear on the EGLE lens is 
\be
\gamma = \sum^{\infty}_{j=1} \frac{\theta_E^2}{(j \Delta)^2}
\, \left(2\pi \Sigma j \Delta^2\right)^{1/2}
= \frac{ \sqrt{2\pi\Sigma} \, \theta_E^2 }{\Delta}\,
\sum^{\infty}_{j=1}\frac{1}{j^{3/2}},
\ee
where $\Delta=1/\sqrt{\pi \Sigma}$ is the dimensionless
distance between two adjacent lenses.
Using $\tau= \pi \Sigma \theta^2_E$, and 
$\sum^{\infty}_{j=1}j^{-3/2}=\zeta(3/2) \simeq 2.6124$
[where $\zeta(x)$ is the Riemann zeta function], we obtain
\be
\gamma = \sqrt{2}\,\, \zeta\left(\frac{3}{2}\right)\, \tau
\simeq 3.7\, \tau.
\ee

\clearpage

\figcaption[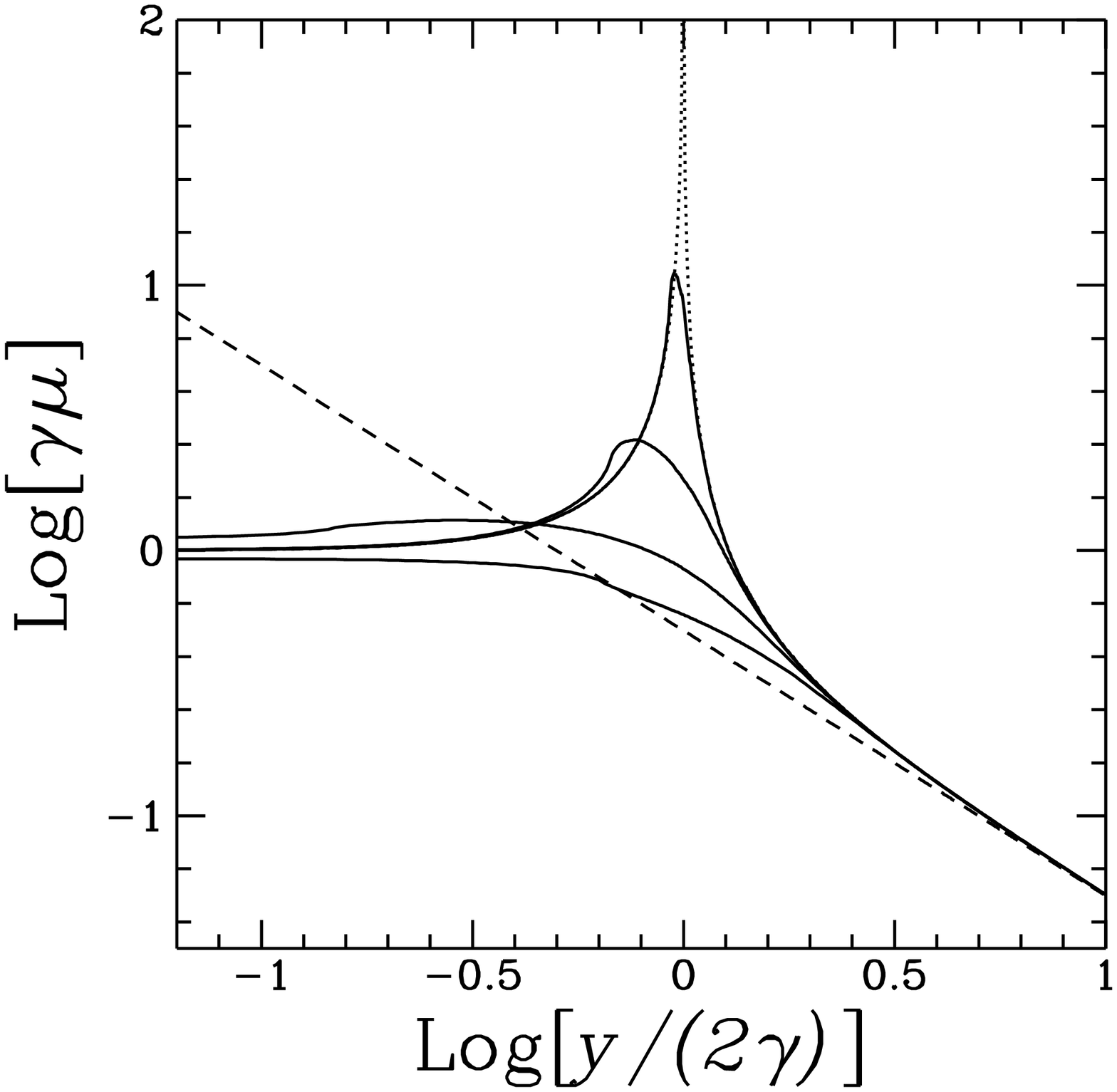]{The lightcurves for $y_2=0$ (cusp-crossing),
$ R/(2\gamma)=0$ (dotted line), 0.01, 0.1, 0.4, 1 (solid lines). 
The dashed line is the lightcurve for $R=\gamma=0$.}
\vspace{0.2in}

\figcaption[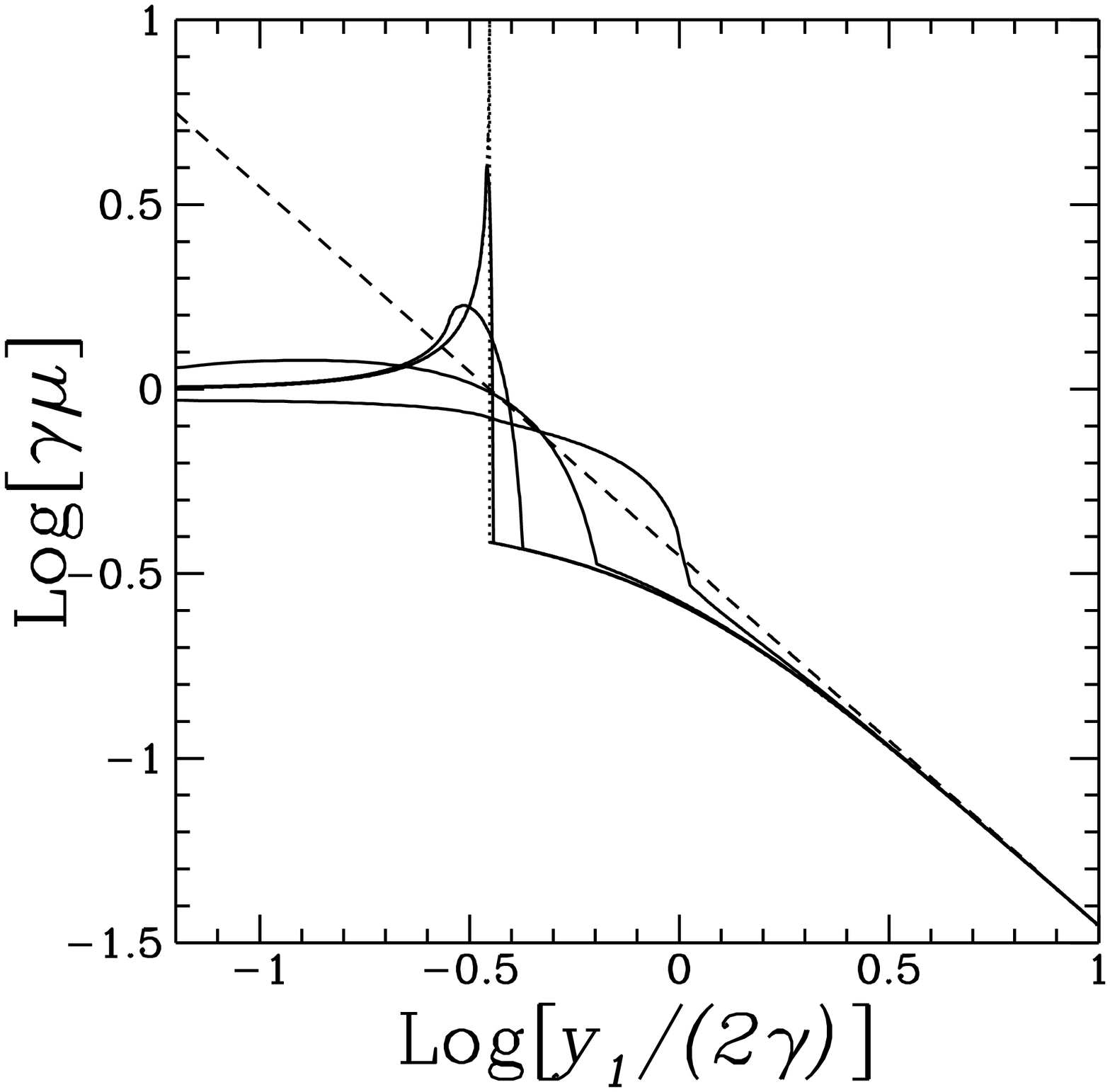]{The lightcurves for $y_2=y_1$ (fold-crossing),
$ R/(2\gamma)=0$ (dotted line), 0.01, 0.1, 0.4, 1 (solid lines). 
The dashed line is the lightcurve for $R=\gamma=0$.}
\vspace{0.2in}

\figcaption[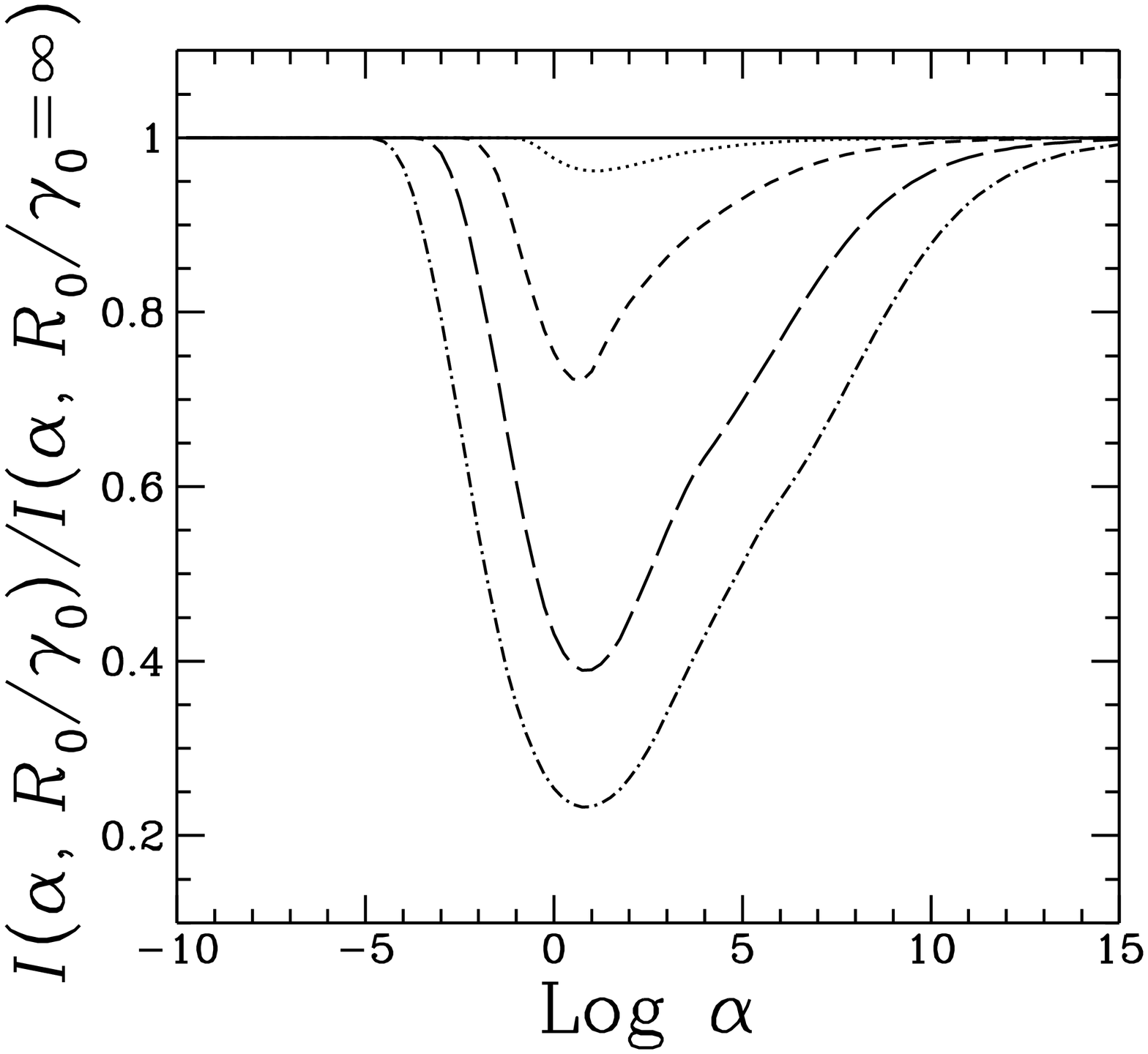]{The modification factor in EGLE volume fraction due to
finite source size and non-zero shear, 
$I_{avg}(\alpha, R_0/\gamma_0)/I(\alpha, R_0/\gamma_0=\infty)$,
for $R_0/\gamma_0=10$ (solid line), 1 (dotted line), 0.1 (short dashed line),
0.01 (long dashed line), 0.001 (dot-short dashed line). }

%\end{document}
\clearpage

\setcounter{figure}{0}
\plotone{fig1.eps}
\figcaption[fig1.eps]{The lightcurves for $y_2=0$ (cusp-crossing),
$ R/(2\gamma)=0$ (dotted line), 0.01, 0.1, 0.4, 1 (solid lines). 
The dashed line is the lightcurve for $R=\gamma=0$.}

\plotone{fig2.eps}
\figcaption{The lightcurves for $y_2=y_1$ (fold-crossing),
$ R/(2\gamma)=0$ (dotted line), 0.01, 0.1, 0.4, 1 (solid lines). 
The dashed line is the lightcurve for $R=\gamma=0$.}

\plotone{fig10.eps}
\figcaption{The modification factor in EGLE volume fraction due to
finite source size and non-zero shear, 
$I_{avg}(\alpha, R_0/\gamma_0)/I(\alpha, R_0/\gamma_0=\infty)$,
for $R_0/\gamma_0=10$ (solid line), 1 (dotted line), 0.1 (short dashed line),
0.01 (long dashed line), 0.001 (dot-short dashed line). }

%\end{document}


\clearpage
\begin{thebibliography}{}
% short labels are from Schneider's book

\bibitem[Blandford and Narayan 1992]{BNrev92}
Blandford, R.D. and Narayan, R. (1992), Ann. Rev. Astron. Astrophys., 30, 311.
 

\bibitem[Chang and Refsdal 1979]{CH79.1}
 Chang, K. and Refsdal, S. (1979), Nature, 282, 561.


\bibitem[Chang 1984]{CH84.1}
 Chang, K (1984), Astr. Ap., 130, 157.

 
\bibitem[Chang and Refsdal 1984]{CH84.2}
 Chang, K. and Refsdal, S. (1984), Astr. Ap., 132, 168.


\bibitem[Schneider and Weiss 1987]{SC87.3}
 Schneider, P. and Weiss, A. (1987), Astr. Ap., 171, 49.


\bibitem[Schneider et al.\ 1992]{Book92}
Schneider, P., Ehlers, J., Falco, E.E., (ed. Springer-Verlag, Berlin, 1992)\/,  ``Gravitational lenses''.

 
\bibitem[Subramanian and Chitre 1985]{SU85.1} 
Subramanian, K. and Chitre, S.M. (1985), \apj, 289, 37.


\bibitem[Wang and Turner 1996]{Wang96}
Wang, Y. and Turner, E. L. (1996), \apj, 464, 114.
\end{thebibliography}
\end{document}